# Conditions for efficient charge generation preceded by energy transfer process in non-fullerene organic solar cells

L. Benatto,[a] C. A. M. Moraes,[a] G. Candiotto,[b] K. R. A. Sousa,[a] J. P. A. Souza,[a] L. S. Roman[a] and M. Koehler[a]

[a] Department of Physics, Federal University of Paraná, 81531-980, Curitiba-PR, Brazil

[b] Institute of Chemistry, Federal University of Rio de Janeiro, 21941-909, Rio de Janeiro-RJ, Brazil

Corresponding authors: lb08@fisica.ufpr.br, koehler@fisica.ufpr.br

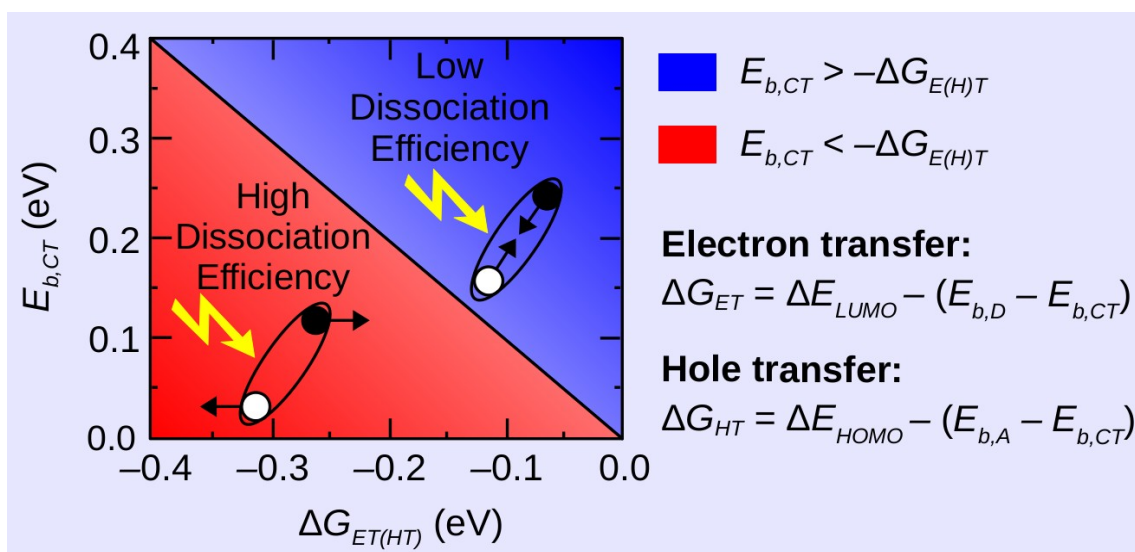

**TOC image**

Abstract

The minimum driving force strategy is applied to promote the exciton dissociation in organic solar cells (OSCs) without significant loss of open-circuit voltage. However, this strategy tends to promote Förster resonance energy transfer (FRET) from the donor to the acceptor (D-A), a consequence generally ignored until recently. In spite of the advances reported on this topic, the correlation between charge-transfer (CT) state binding energy and driving force remains unclear, especially in the presence of D-A FRET. To address this question, we employ a kinetic approach to model the charge separation in ten different D/A blends using non-fullerene acceptors. The model considers the influence of FRET on photoluminescence (PL) quenching efficiency. It successfully predicts the measured PL quenching efficiency for D or A photoexcitation in those blends, including the ones for which the D-A FRET process is relevant. Furthermore, the application of the model allows to quantifying the fractions of quenching loss associated with charge transfer and energy transfer. Fundamental relationships that controls the exciton dissociation was derived evidencing the key roles played by the Marcus inverted regime, exciton lifetime and mainly by the correlation between the driving force and binding energy of CT state. Based on those findings, we propose some strategies to maximize the quenching efficiency and minimize energy loss of OSCs in the presence of D-A FRET.

# 1. Introduction

The singlet excitons ($S_1$), formed after the photoexcitation of organic semiconductors, are not easily dissociated because the electron-hole pairs have binding energy ($E_b$) much higher than thermal energy at room temperature.[1] This scenario stems from the relatively low dielectric constants (ε ≈ 2-5) of organic semiconductors.[2] Therefore, for the application of these materials in organic solar cells (OSCs) one additional energy is necessary to promote the exciton dissociation.[3] Bulk heterojunction (BHJ) active layers combining electron donor (D) and acceptor (A) components can provide the additional energy needed for the dissociation of photogenerated excitons, which involves the creation and dissociation of an intermediary charge transfer (CT) state to form a final charge-separated (CS) state.[4] If the photon is originally absorbed in the donor material (and the Förster resonance energy transfer (FRET) from D to the A is negligible), the electron transfer from D to A is the first step for exciton dissociation. Considering the excitation of acceptor material, the first step is now a hole transfer from A to D.[5] Although an efficient exciton dissociation has already been obtained in several D/A blends, a strategy to further increase the open-circuit voltage ($V_{OC}$) of OSCs involves the combination of D and A materials with the minimal driving force to promote the exciton splitting.[6–9] This design strategy, however, tends to favor the FRET process from donor to the acceptor.[10–12] Some works suggest that hole or electron wave function delocalization can be fundamental to provide the exciton dissociation with a low driving force.[13–17] Yet the mechanism behind the production of free charges in OSCs is still unclear.[18–22] For instance, the presence of the FRET from D to A tends to decrease the relative contribution of electrons transferred from D to A in favor of holes transferred from A to D as the main channel of exciton deactivation. To clarify these issues, more in-depth studies of the charge transfer process on D/A interfaces with different driving forces must be conducted.[23–26]

The fine-tuning of energy level alignment in BHJ is now possible with the development of new non-fullerene acceptors (NFAs).[27–29] The flexible molecular structure of NFAs allows simple design modifications,[30–32] like the addition of fluorine or chlorine in the chemical structure,[33,34] which enables an increase of the highest occupied molecular orbital (HOMO) level and lowest unoccupied molecular orbital (LUMO) level. In addition, the optical gap ($E_{opt}$) and $E_b$ are also sensitive to design modifications of NFAs.[35–38] Therefore, a systematic investigation with experiments and simulations using different D/A combinations (including a variety of NFAs) is essential to reveal the conditions that the driving force must fulfill in order to promote an efficient exciton dissociation.[39] This was performed recently by Yang *et al.* in a seminal manuscript.[40] They selected a particular polymer, PBDB-TF[41] (also known as PM6 and PBDB-T-2F[42]) and produced OSCs by blending it with ITIC,[43] IEICO[44] and its derivatives (ITCC,[45] IT-M,[46] IT-2F,[47] IT-2Cl,[48] IT-4F,[49] IT-4Cl,[48] IEICO-4F[50] and IEICO-4Cl[51]), studying a total of 10 D/A combinations. A careful investigation of the charge separation process was performed by them with a series of experiments, including photoluminescence (PL) measurements. The authors point out that a certain energy-level offset between D/A is necessary to provide sufficient driving force for free charge generation, but they were unable to establish a threshold equally valid for photoexcitation of the donor and acceptor materials. As a suggestion, the authors propose an in-depth analysis *via* a theoretical calculation to define the conditions for efficient exciton dissociation and its relationship with molecular structure.

The comparison between the PL of pristine D or A materials with the PL of D/A blend (involving both experiment and theory) has been the standard procedure over the years to study the exciton dissociation process.[52–54] A complete quenching of the excitonic luminescence after D or A excitation indicates that all excitons are dissociated in the blend. Therefore, by measuring the PL quenching efficiency, the effects of energy-level alignment in the exciton dissociation process can be studied. Fortunately, Yang *et al.* performed these measurements for the 10 blends studied, selectively exciting D or A and providing 20 PL quenching efficiency measurements. In this work, we will generalize the kinetic model for PL quenching calculation[55,56] to consider the FRET phenomena. We will show that a good theoretical description of the experimental results obtained by Yang *et al.* is possible only after considering the FRET transfer between donor and acceptor. It was recently demonstrated that excitons transferred *via* FRET from the donor to the NFA can be an important deactivation channel of the donor-excited state.[10,57]

The basic idea here is to systematically confront theoretical results with experimental data to reveal the properties of the driving force necessary to promote efficient exciton dissociation. An important parameter of the kinetic model is the $E_b$ magnitude of each material. Thus, we propose a mixed approach to calculate the dielectric constant of the organic film that involves density functional theory (DFT) calculations combined with classical molecular dynamics (MD) to simulate solvent evaporation to form the organic film. The combination of DFT with MD is useful to study organic materials.[58–61] Following this procedure a reliable estimate of the solid-state $E_b$ can be obtained. We will then use those results to obtain the driving forces and Marcus rates for the processes involved in the dynamics of the electron and hole at the D/A interface. Applying the rates in an analytical expression, considering disorder effects, it will be possible to obtain the PL quenching efficiency and to study its variation with some important parameters of the model. The application of our model to the data published in ref.[40] established an important condition that the electronic structure of the D/A system must fulfill in order to guarantee an efficient exciton dissociation. Particularly, the model revealed the fundamental contributions of the Marcus inverted regime, the exciton binding energy and exciton lifetime to enhance the charge generation process. This result can be an important guideline to the fabrication of improved OSCs.

## 2. Calculation details
### 2.1 Exciton binding energy

The first step to calculate $E_b$ is to optimize the ground state geometry of isolated materials by DFT with the B3LYP[62] functional and 6-31G(d,p) basis set. Based on the optimized geometries the molecular energies will be calculated by DFT/TDDFT with the long-range-corrected hybrid functional $\omega$B97XD[63] and the 6-31G(d,p) basis set.[64] This calculation was done considering an optimized range-separation parameter ($\omega$). The $\omega$ value was optimized by minimizing the expression $J(\omega)$:[65]

$$J(\omega)=|E_{HOMO}(\omega)-IP(\omega)|+|E_{LUMO}(\omega)-EA(\omega)| \qquad (1)$$

where $E_{HOMO}(\omega)$ and $E_{LUMO}(\omega)$ are the energies of the HOMO and LUMO (respectively, highest occupied and lowest unoccupied molecular orbitals), while IP($\omega$) and EA($\omega$) are the vertical first ionization potential and electron affinity of the material, respectively. With the total energies of the cationic ($E_+$), anionic ($E_-$), and neutral ($E_0$) states, it is possible to calculate IP = $E_+ - E_0$ and EA = $E_0 - E_-$.[66] The $E_b$ can be estimated by subtracting the fundamental energy gap ($E_{fund}$ = IP - EA) from the $E_{opt}$. The $E_{opt}$ can be obtained from TDDFT calculations and corresponds to the energy of the first optically allowed excited state ($S_1$). We also calculated the atomic charges *via* the electrostatic potential (ESP) method.[67,68] All DFT/TDDFT calculations were performed using the Gaussian 16 package.[69]

**2.2 Dielectric constant**

To consider the effect of the surrounding dielectric medium (electronic polarization effect) on exciton binding energy, it is necessary to obtain the dielectric constant ($\varepsilon$) of materials. Just for the polymer, we will use an $\varepsilon$ of 4 in the calculations, in accordance to previus works.[56,70,71] The $\varepsilon$ value for the acceptor molecules can be estimated through Clausius Mossotti equation for nonpolar materials, as described below:[72]

$$\frac{\varepsilon-1}{\varepsilon+2} = \frac{4\pi}{3} \frac{\rho}{M} N_A \alpha, \qquad (2)$$

where $\rho$, $M$, $N_A$, and $\alpha$ are the density of the material, molecular mass, Avogadro number, and the isotropic component of the molecular polarizability, respectively.[73] The $\rho N_A/M$ is the reciprocal of the molecular volume ($V_M$). Most works obtain $V_M$ and $\alpha$ from the neutral molecular geometry optimized by DFT (tight option is generally used in the self-consistent field (SCF) convergence for better accuracy of $V_M$).[15,35,72–74] However this method neglect geometry changes of the molecules induced by clustering effects in organic thin films.[60,61,75,76] In our work, we implement a more robust method to obtain $V_M$ and $\alpha$ combining DFT with MD simulations. We first simulate the energy minimization and thermodynamic equilibrium of a molecular cluster considering 50 acceptor molecules in chlorobenzene solvent *via* MD (using GROMACS package[77]) The acceptor molecules were randomly inserted in a cubic box and the empty spaces were filled with the previously thermalized chlorobenzene solvent (Fig. S1) (details in Supporting Information (SI)). After the solvent evaporation process, molecular clusters were formed (Fig. S2). Thus, it is possible to obtain $V_M$ through the Fractional Free Volume (FFV)[78] calculated by GROMACS[79] (full description in SI). In the calculation of $\varepsilon$, we consider α being the average value calculated for the 50 individual molecules. Using this procedure, it is expected to obtain a more reliable value for dielectric constant. Finally, the solid-state exciton binding energy is obtained such as $E_b^{vac}/\varepsilon$. In our previous work,[15] we found that $E_b^{vac}/\varepsilon$ produce similar results compared to the results obtained by the polarizable continuum model.[80]

**2.3 PL quenching efficiency**

Recently, it was suggested that FRET from the polymer to NFA is the main deactivation process of the donor photoexcited state.[10,57] Indeed there is considerable spectral overlap of the PBDB-TF emission and NFA absorption for molecules considered in ref.[40]. Yet, this condition is not satisfied between the PL of the acceptor and the absorption of the donor. Based on these findings we generalized the kinetic model proposed in ref.[55] to consider the FRET process between donor and acceptor.

The model is based on a simple one-electron picture, describing the electron and hole kinetics at the D/A interface by a sequence of processes that are characterized by a frequency rate ($k$). In Fig. 1, we illustrate the charge and energy transfer dynamics considered here, with the rates assumed to describe the PL quenching efficiency via donor or acceptor excitation. After donor excitation, there is competition between FRET and charge transfer to produce PL quenching. Considering the photoexcited donor it can either: (i) transfer its energy to another donor (with a rate $k_{F,DD}$); (ii) transfer its energy to a nearby acceptor (with a rate $k_{F,DA}$); (iii) transfer its electron to the acceptor (with a rate $k_{ET}$); or, (iv) finally, the singlet exciton will recombine (with a characteristic rate $k_{SR,D}$). We assume that the occurrence of those phenomena is mediated by a probability $p_{F,DA}$. In essence $p_{F,DA}$ depends on the number of excited donors that transfer their energy to the acceptor in an infinitesimal time interval $dt$ (for more details see the supporting information section S2). When $p_{F,DA}$ is high, the FRET process prevails over the electron transfer as a channel to deactivate the population of excitons in the donor (we will discuss more about $p_{F,DA}$ below).

Once the energy is transferred by FRET to the acceptor the excited NFA can either transfer the hole to the donor (with a rate $k_{HT}$) or recombine (with a characteristic rate $k_{SR,A}$) (see Fig. 1 again). The presence of FRET thus links the exciton quenching for donor excitation ($Q_D$) with the kinetics of hole transfer to the acceptor. In the supporting information section we developed all mathematical steps followed to write the $Q_D$, assuming the steady state approximation:

$$Q_D = 1 - k_{SR,D} f_D \left( (1 - p_{F,DA}) + p_{F,DA} k_{F,DA} f_A \right). \tag{3}$$

where

$$f_D = \frac{k_{EB} + k_{ER} + k_{ES}}{(k_{SR,D} + k_{ET} + p_{F,DA} k_{F,DA})(k_{EB} + k_{ER} + k_{ES}) - k_{EB} k_{ET}}. \tag{4}$$

The $f_A$ in Eq. 3 will be defined in Eq. (7) below. The subscripts ET, EB, ER, and ES represent the electronic processes of transfer (from $S_{1,D}$ to CT), back (from CT to $S_{1,D}$), recombination (from CT to $S_0$), and separation (from CT to CS, which involves the transfer of electrons from the nearest acceptor molecule adjacent to the polymer to the next-to-the nearest acceptor molecule), respectively. $S_{1,D}$ refers to the singlet exciton in the donor, CT the charge transfer exciton at the D/A heterojunction and $S_0$ the fundamental state of the donor.

Using Eq. (3), it is possible to define $f_{CT} \equiv (1 - p_{F,DA}) k_{SR,D} f_D$, which corresponds to the fraction of quenching loss associated with electron transfer, and $f_F \equiv p_{F,DA} k_{SR,D} k_{F,DA} f_D f_A$, which is the fraction of loss associated with the FRET-hole transfer processes. Those quantities are useful since they allow to quantifying the relative contribution of each deactivation channel in the total quenching loss.

The quenching dynamics described by equation (3) will be completely determined once the probability $p_{F,DA}$ is known. We will assume that $p_{F,DA}$ depends on the process that deactivates (or reactivates) the singlet donor state.[81] This probability will also depend on the average density of donor ($n_D$) or acceptor ($n_A$) molecules surrounding a determined excited donor that are available to receive the exciton, *i.e.*, the number of acceptors per donor ($n_D/n_A$).[81] For a homogeneous blend (in which donor and acceptor have a similar average density) $n_A/n_D \approx 1$. Additionally, the D/A ratio of the ten blends considered in this work is 1:1 (weight by weight, w/w).[40] Under those assumptions $p_{F,DA}$ is given by the $k_{F,DD}$ rate divided by the total rate that describes all the processes that can decrease the concentration of $S_{1,D}$, or

$$p_{F,DA}=\frac{k_{F,DA}}{k_{F,DA}+k_{F,DD}+k_{SR,D}+k_{ET}-k_{EB}}. \tag{5}$$

For example, when $k_{ET}$ (or $k_{F,DD}$) is much higher than $k_{F,DA}$, $p_{F,DA} \approx 0$ and the influence of the energy transfer from D to A on exciton dissociation is negligible. In this case, the expression for $Q_D$ in Eq. (3) is reduced to the formula proposed in ref.[55] (deduced without considering the FRET between donor and acceptor). Curiously, if the hole transfer from A to D is sufficiently high, the increase in $Q_D$ would be obtained by a poor electron injection (that will increase $p_{F,DA}$). This result is counterintuitive.

For the selective light excitation of the acceptor, the energy transfer between the acceptor and donor is absent. Therefore, the equation to obtain the PL quenching efficiency in steady-state approximation for acceptor excitation ($Q_A$) is equal to our previous work[55]

$$Q_A=1-k_{SR,A}f_A, \tag{6}$$

where

$$f_A=\frac{k_{HB}+k_{HR}+k_{HS}}{(k_{SR,A}+k_{HT})(k_{HB}+k_{HR}+k_{HS})-k_{HB}k_{HT}}. \tag{7}$$

The subscripts HT, HB, HR, and HS represent the hole transfer (from $S_{1,A}$ to CT), back (from CT to $S_{1,A}$), recombination (from CT to $S_0$) and separation (from CT to CS, which involves transferring a hole from the nearest donor polymer adjacent to the acceptor molecule to the next-to-the nearest donor polymer), respectively. The subscript SR represents singlet exciton recombination. $k_{SR,A}$ is the inverse of singlet exciton recombination lifetime in the acceptor and can be obtained from experimental data. In view of the structural similarity between the acceptors and the lack of measurements for the materials considered, the $k_{SR,A}$ of all acceptors is derived from the recombination lifetime of singlet excitons in ITIC.[82]

**2.4 Transfer rates**

Fluorescence (or Förster) resonance energy transfer is a nonradiative energy transfer mechanism based on dipole-dipole coupling, where a donor molecule in an electronically excited state transfers its excitation energy to a nearby acceptor molecule. For efficient energy transfer, it is necessary that the emission spectrum of the donor molecules overlaps the absorption spectrum of the acceptor molecules, and the separation distance between the donor and acceptor centers has to be much less than the wavelength.[83] In FRET model, the FRET rate $k_F$ between donor and acceptor is given by ref[84]

$$k_F = \frac{1}{\tau_D}\left(\frac{R_0}{R}\right)^6 \qquad (8)$$

where $R_0$ is the Förster radius, $R$ is the donor-acceptor distance and $\tau_D$ is the exciton lifetime of the donor in the absence of the acceptor (equal to 178 ps for PBDB-TF[10]). The procedure for calculating $R_0$ is well known in the literature and is described in the supplementary information (see Figs. S3-S6 and Table S3). Figure S6 shows the high overlap between the emission spectrum of the PBDB-TF polymer and the absorption spectrum of the ITIC derivatives.

We used here the Marcus/Hush equation[85] to obtain the characteristic frequency of the rates involved in the exciton dissociation process

$$k = \frac{4\pi^2}{h}\frac{\beta^2}{\sqrt{4\pi\lambda k_B T}}\exp\left[\frac{-\Delta G^{\ddagger}}{k_B T}\right] \quad, \qquad (9)$$

where $k_B$, $T$, $\lambda$, and $\beta$ are the Boltzmann constant, temperature, reorganization energy, and electronic coupling (transfer integral). The inner component of the reorganization energy was estimated using the adiabatic potential energy surfaces of the neutral and charged molecules.[55] The outer component of the reorganization energy, much lower compared to the internal component,[86] was set as 36 meV, a physically plausible parameter for the external component.[87] The activation energy for charge transfer, $\Delta G^{\ddagger}$, is given by the folowing[88]

$$\Delta G^{\ddagger} = \frac{(\lambda + \Delta G)^2}{4\lambda} \quad, \qquad (10)$$

where $\Delta G$ is the Gibbs free energy (driving force) of the charge transfer reaction, approximated here as the energy difference between initial and final states. The rate given by Eqs. (9) and (10) increases when $\Delta G \rightarrow \lambda$ and has a maximum at $\Delta G = \lambda$. A negative value for $\Delta G$ indicates that the final state has lower energy than the initial state, *i.e.*, the process is thermodynamically favorable.

**2.5 Driving forces**

Fig. 2a illustrates the energies involved on charge dynamics based on the states energies description.[23] Fig. 2b and c summarize all definitions of driving forces involved in the electron (hole) dynamics at the D/A interface[70,89] as derived from the description of Fig. 2a. The energies $E_{LE}$, $E_{CT}$ and $E_{CS}$ are the local excited-state energy (optical gap), the charge transfer energy and the charge separated energy, respectively. A rapid inspection to Fig 2a shows that this driving force for electron transfer, $E_{CT} - E_{LE,D}$, can be expressed as a function of exciton binding energies or $\Delta G_{ET} = \Delta E_{LUMO} - (E_{b,D} - E_{b,CT})$, where $E_{b,CT}$ is the binding energy of the CT exciton. Likewise, the corresponding driving force for hole is $\Delta G_{HT} = \Delta E_{HOMO} - (E_{b,A} - E_{b,CT})$. It is often assumed that the strong dielectric stabilization of the two charges forming a CT state compensates the coulomb binding of a Frenkel exciton in the solid state, making $E_{b,A} \approx E_{b,CT}$ and $\Delta G_{HT} \approx \Delta E_{HOMO}$.[10] We will test this hypothesis in section 3.

In several works, $\Delta E_{LUMO}$ and $\Delta E_{HOMO}$ are measured by cyclic voltammetry (CV)[46,90] or ultraviolet photoelectron spectroscopy (UPS).[49,91] We will use CV measurements through by Yang and coworkers[40] to obtain those energies for the ten D/A blends under consideration. In the same line, we will use the simple approximation $E_{CT} = E_{LUMO,A} - E_{HOMO,D}$ to obtain $\Delta G_{ER}$ and $\Delta G_{HR}$.[92,93] In Table S4, we compare $E_{CT}$ obtained from CV and sophisticated electroluminescence spectra measurements[10,94] for three of the ten blends studied here. The deviation between the energies presented in Table S4 was 0.09, 0.05 and 0.02 eV, with the average deviation was low, approximately 0.05 eV. Since we consider that the characteristic energies can vary due to disorder (more details will be presented below), the effects of these energy differences are averaged out during the calculations. Therefore, we consider that obtaining the energies through CV does not significantly affect the results of the simulations.

## 2.6 The effect of disorder

Due to the amorphous nature of organic semiconductors, the characteristic energies and parameters for charge transfer can fluctuate due to the presence of the diagonal and off-diagonal disorder.[85] Consequently, a realistic theoretical description of the charge transfer dynamics at the D/A interface must consider these two kinds of disorder effects. The diagonal disorder reflects the fluctuations of the energy levels (driving forces). To consider this effect, a random value from a Gaussian distribution centered around zero with standard deviation $\sigma = 0.1$ eV will be added to the value of each driving force defined in Figs 2b and c. This magnitude of $\sigma$ is typical of diagonal disorder fluctuations in organic semiconductors, as discussed in previous works.[95–97]

The off-diagonal disorder is related to fluctuations in the electronic coupling between adjacent molecules. It can thus influence the transfer integral factor $\beta$ in the Marcus/Hush equation (Eq. (9)). For calculating all the rates, we will define $\beta = \beta_{max} \cos(\varphi)$, where $\beta_{max}$ is the maximum value of electronic coupling and $\varphi$ is a random angle between 0 and $\pi/2$. In this way, we are considering adjacent molecules with different shapes between face-on configuration ($\varphi=0$ and $\beta = \beta_{max}$) and edge-on configuration ($\varphi=\pi/2$ and $\beta = 0$). The $\beta_{max}$ will be set as 50 meV for all the processes, a typical value of electronic coupling for organic semiconductors in face-on configuration.[37,97–99] Considering this approach to obtain $\beta$, we are also limiting the differences between the blends to variations in $\lambda$, $E_b$, and frontier energy levels. This is justified by the fact that Yang and coworkers showed similar surface morphologies, crystallinity, and molecular packing for all blends, explained by the similar chemical structures of these NFAs.

When disorder is considered, the PL quenching efficiency calculated using Eqs (3) and (6) becomes unique for each exciton. It is then necessary to calculate average quantities to characterize the charge transfer process for a particular D/A system. To obtain numerical values with sufficient precision, all parameters are averaged over $10^4$ runs of the simulation.[56,95] Following this procedure, the average exciton quenching efficiency for the donor and acceptor excitation is computed. It is important to highlight then that the results reported below for quenching efficiencies and charge transfer rates are averaged values.

## 3. Results and Discussions

### 3.1 Correlation between theory and experiment

#### 3.1.1 Dielectric constant

For each material, the optimal range-separation parameter, $\omega$, and the molecular energies to obtain $E_b^{vac}$ are presented in Table S5. In Table 1, we can see that IEICO-based molecules present the lowest values of $E_b^{vac}$. This result is related to its greater molecular length, which facilitates the electron–hole pair delocalization.[100] Note that among acceptors, the smallest molecular length of ITCC causes the largest $E_b^{vac}$. Considering separately the ITIC and IEICO derivatives, the inclusion of fluorine and mainly chlorine in the end groups is beneficial to decrease $E_b^{vac}$. This modification increases the internal charge transfer (ICT) of acceptor molecules, which induces a smaller overlap between the electron–hole pair.[15,50] The PBDB-TF copolymer showed a higher $E_b^{vac}$ that can be explained by its lower ICT character compared to the A–D–A-type aromatic hybrid fused-ring electron acceptors. In Fig. 3 we present the partial atomic charges for the simulated materials in the ground state geometry.

Table 1 also shows the $\varepsilon$ value obtained for each material and the parameters used for its calculation. Molecular volume was calculated from the routine of GROMACS package and using the Eq. (S3) for molecular clusters containing 50 acceptors (Fig. S2). The magnitude of isotropic polarizability for 50 acceptors can be seen in Fig. 4 and the average in Table 1. The IEICO derivatives presented the higher magnitudes of molecular volume and mean isotropic polarizability. Applying these two quantities in the Clausius Mossotti equation the dielectric constant was obtained. There is a good agreement between theory and experiment for the dielectric constant of ITIC molecule. The IEICO based molecules have a high dielectric constant compared to the other NFAs in Table 1. This feature helps to decrease the solid-state exciton binding energy ($E_b^{vac}/\varepsilon$) of the IEICO-based molecules, which was found to be in the interval 0.33-0.34 eV for these acceptors. Among the ITIC derivatives, IT-4Cl presented the lower $E_b^{vac}/\varepsilon$, equal to 0.37 eV. The magnitude of solid-state exciton binding energy for each material will be important for calculating the driving force for charge transfer.

### 3.1.2 Exciton Quenching

As mentioned earlier, in the solid state, the strong dielectric stabilization of two charges in a CT state compensates the coulomb binding of a Frenkel exciton, making $E_{b,A} \approx E_{b,CT}$ and $\Delta G_{HT} \approx \Delta E_{HOMO}$. We tested this assumption in our calculations of $Q_A$. We found that there is a poor agreement between the theoretical values of $Q_A$ and the experimental results obtained by Yang *et al.* (see Fig. S7a) when using this approximation. The adjusted squared correlation coefficients ($R_{sq}$) obtained from linear fitting was only 0.84 (the quality of the linear regression to reproduce the data improves as $R_{sq}$ approaches 1). Alternatively, there is a considerable improvement of the theoretical description if one slightly adjusts the $E_{b,CT}$ in relation to $E_{b,A}$ (an adjustment not greater than 0.1 eV was required) as can be seen in Table 1. Following this procedure, there is an excellent agreement between the theoretical value of $Q_A$ and the experiment, as can be seen in Fig. 5a, $R_{sq}$ is near 1.

With the determination of $E_{b,CT}$, it then possible to obtain $Q_D$. Yet, we found that a precise description of $Q_D$ is impossible if the energy transfer between the donor and the acceptor is neglected (see Fig. S7b). Note that in this case theoretical quenching ($Q_D$ – Theo.) are higher than the experimental one ($Q_D$ – Exp.) for some blends. To correctly describe $Q_D$ for blends where $k_{F,DA}$ is relevant, acceptor quenching efficiency must be considered upon selective excitation of the donor.[101] When FRET is present and $Q_A$ is not close to 100%, the acceptor PL can be detected after the donor photoexcitation.[10] In fact, this effect was observed in the PL measurements performed by Yang *et al.* for the blends involving ITIC derivatives.[40] The good agreement between the theory developed in Eq. (3) and the experimental $Q_D$ can be seen in Fig. 5b. The results demonstrate that the FRET mechanism is relevant for blends based on ITIC derivatives.

To complement the results of Fig. 5a and b, Fig. 6a and b show the driving forces for hole and electron dynamics, respectively. Additionally, all the rates involved in the calculation of PL quenching efficiency are presented in Fig. 6c and d (averaged over $10^4$ runs of the simulation, see details in section 2.6). The Marcus rates are influenced not only by the respective driving forces but also by the reorganization energies.[102] In Table S6 the reorganization energies were found to be around 0.23-0.35 eV. To facilitate the observation of the results shown in Fig. 6, we present in the Tables S7 to S9 of supplementary information all the numerical values of the driving forces and the respective Marcus rate.

Due to negative values of $\Delta G_{ET}$ in Fig. 6b, the electron transfer rate in Fig. 6d is the processes with higher values compared to the other rates (excluding the FRET rate). Note in this figure that the electron transfer rate, $k_{E,T}$, has a magnitude close to the FRET rate between donors, $k_{F,DD}$, and FRET rate between donor and acceptor, $k_{F,DA}$. This shows that there is indeed a competition between the FRET and the charge transfer as channels for excitation dissociation, especially for the ITIC derivatives. For this family of NFAs $k_{F,DA} \approx k_{ET}$, which helps raise the FRET probability $p_{F,DA}$ (see Eq. 5 and Table 2). The opposite happens to the IEICO family. For those acceptors $k_{ET}$ and $k_{F,DD}$ are higher than $k_{F,DA}$, which decreases $p_{F,DA}$. This essentially happens because there is a lower overlap between the polymer photoluminescence and the acceptor absorption for those NFAs (see Figure S6).

It is possible to see in Fig. 6c that the hole transfer rate, $k_{HT}$, has the highest value for most blends. Only for the blends with IEICO derivatives that this rate is lower than the hole back rate, $k_{HB}$, because of the unfavorable driving force for hole transfer. Additionally, the hole separation rate, $k_{HS}$, has a lower frequency than $k_{HB}$ for those acceptors. The combination of these two factors reduces $Q_A$, as seen in Fig. 5a. These results suggest that the exciton dissociation in the blends using IEICO derivatives basically relies on the efficient electron transfer from the donor to the acceptor. In this way, the mechanism for PL quenching in blends involving PBDB-TF polymer with those NFAs resembles the mechanism found in traditional fullerene acceptors. Considering the PBDB-TF/ITCC blend, it is observed in Fig. 6c that the rate for hole separation has a lower frequency than the rates for hole back and singlet recombination on acceptor, $k_{SR,A}$. This fact decreases the efficiency of PL quenching for acceptor excitation.

Because of the high presence of FRET, the process of donor exciton quenching for blends containing the ITIC family of NFAs majority depends on the hole transfer from the acceptor to the donor. The highest efficiency of exciton quenching was obtained for the blends using halogenated derivatives of ITIC. Those systems also have the highest values of FRET probability, $p_{F,DA}$. The enhancement of this probability is not essentially derived from a higher $k_{F,DA}$ since there are molecules that presented higher frequencies for this rate (see Table S3). Indeed, this rate tends to decrease slightly with the insertion of the F or Cl atoms because the resulting planarization of the molecules shifts the absorption to higher wavelengths (thus decreasing the overlap with the PL of the polymer). The high values of $p_{F,DA}$ for those blends rests basically on the fact that they have a slightly lower value of $k_{ET}$. This happens because the Marcus rate of electron transfer in those systems is in the so-called Marcus inverted region (MIR) so that $|\Delta G| > \lambda$. In this regime a more negative activation energy induces a lower rate for charge transfer. Consequently, the main channel for exciton deactivation after donor excitation is then a multiple step process that involves the exciton migration from donor to acceptor and the following hole transfer to the donor (with the formation of the CT state). Since those blends have an efficient hole transfer, both $Q_D$ and $Q_A$ are high in those systems.

In this work, the non-radiative (NR) recombination rates from CT ($k_{HR}$ and $k_{ER}$) were modeled using the Marcus/Hush equation. From Fig. 6c-d, the $k_{HR}$ and $k_{ER}$ rates are lower than the other rates associated with the transfer and separation processes. Consequently, the CT recombination rate weakly influences the magnitudes of $Q_D$ and $Q_A$. In a recent report,[97] the NR recombination rate *via* CT state was calculated for a range of $E_{CT}$ values by using the Marcus–Levich–Jortner model that considers the thermal population of high frequency vibrational modes.[103] This stimulated us to recalculate the quenching efficiencies using these interesting results. It is possible to see in Figure SX that no significant variations in quenching efficiencies were observed with this procedure, demonstrating that high frequency behavior is less important for the systems studied here. In another recent work,[10] the NR recombination rate from CT to the ground state was estimated to be around $10^8$ s$^{-1}$ for blends with high $E_{CT}$ values, in agreement with our results obtained using Marcus/Hush equation. Note that the D/A blends studied in our work have high $E_{CT}$ values, which minimizes the NR recombination rate.[70,104]

In concluding this section, although Fig. 6a-d helped to interpret the results of Fig. 5a-b, we intend to find a simpler way to understand the exciton dissociation process just by looking at the state energies (see the later section 3.3.1).

## 3.2 Relative contribution of FRET and electron transfer to quenching loss

Table 2 compares the FRET probability, $p_{F,DA}$, the contributions of electron transfer, $\eta_{CT}$, and FRET-hole transfer, $\eta_F$, to the total quenching loss, and the quenching efficiency for acceptor excitation, $Q_A$. $\eta_{CT(F)}$ is calculated using $f_{CT}$ and $f_F$ from Eq. 3 so that $\eta_{CT(F)} = f_{CT(F)}/(f_{CT} + f_F)$. One can see that the blends formed by the acceptors from IT-2F to ITIC are characterized by a strong contribution of the FRET channel to the loss of quenching ($\eta_F > 99\%$). But the origin of such a high contribution is slightly different among those acceptors: whereas the IT-nX family (where n = 2 or 4 and X is F or Cl) have $Q_A > 90\%$ and $p_{F,DA}$ usually above 60% (specially for the chlorinated derivatives), the ITCC, IT-M and ITIC have $59\% < Q_A < 85\%$ and lower $p_{F,DA}$. Consequently, the high $\eta_F$ in the IT-nX group comes simply from the fact that most excitons generated in the donor are transmitted to the acceptor. Most part of those excitons will efficiently dissociate by hole transfer to D, but some will recombine and decrease the quenching. For ITCC, IT-M and ITIC, however, less excitons are transmitted by FRET. Despite the lower number of excitons, the poor efficiency of hole transfer for those systems will keep $\eta_F$ around 99% due to improved recombination. Considering now the IEICO family, $p_{F,DA}$ is below 10% for the blends using those acceptors. Hence, most of the excitons are dissociated by electron transfer from donor to acceptor. Although $\eta_{CT}$ considerably increases for this group, a substantial fraction of the quenching loss is still associated with FRET transmission to the acceptor. This happens because the hole transfer is inefficient (which can be verified by the low $Q_A$). Once the excitons reach the acceptor, they will tend to recombine.

Interestingly, there is an apparent correlation between values of $p_{F,DA}$ and $Q_A$ in Table 2. This suggests a relationship between those parameters that in principle could be applied to optimize the performance of the OSC in the presence of FRET. We will investigate this possibility in section 3.3.2.

## 3.3 Strategies to maximize the quenching efficiency
### 3.3.1 Correlation between driving forces and exciton binding energy

The good agreement between the model predictions and the experiment motivated us to use our theory as a tool to explore the relative influence of the system's characteristic energies at the D/A heterojunction on the magnitudes $Q_D$ and $Q_A$. This can be an intricate task due to the peculiarities of the D/A system. For instance, recently Nakano *et al.*[23] studied the correlation of state energies with charge generation efficiency of 16 combinations of four donor polymers and four acceptor molecules in planar heterojunction. They surprisingly found that the electron-hole binding energy in the CT state (same as driving force for charge separation $\Delta G_{ES}$ and $\Delta G_{HS}$ as shown in Fig. 2) is not a key factor for free charge generation. Our results lead to the same conclusion as Nakano, see Fig. S9.

We would like to begin our discussion by looking at the correlation between the $\Delta E_{LUMO(HOMO)}$ and $Q_{D(A)}$. Fig. 7a shows that when $\Delta E_{LUMO(HOMO)}$ approaches zero, $Q_{D(A)}$ became smaller. Thus, negative values of $\Delta E_{LUMO(HOMO)}$ remain important to promote the dissociation of excitons even when the D-A FRET is present. The same is true when considering $\Delta G_{ET(HT)}$ instead of $\Delta E_{LUMO(HOMO)}$, see Fig. 7b. From those figures, $Q_D$ begins to fall earlier that $Q_A$, with an energy difference greater than 0.1 eV. In Figs 7a and 7b, we fitted the experimental data using a second order polynomial. This function can reproduce separately the experimental quenching for the donor and acceptor excitation. However, there is no single set of parameters that can fit simultaneously $Q_D$ and $Q_A$. This indicates that, although these energies may be useful to optimize the quenching, they do not constitute a fundamental parameter to anticipate quenching efficiency.

A simple criterion to characterize systems with a high quenching efficiency (derived exclusively from charge transfer) was proposed by us in ref.[55]. It was suggested that, after the CT formation, exciton dissociation *via* electron or hole transfer will mainly involve the competition between the charge back process (to recreate the singlet state) and the process that dissociates the CT state into free charges. This reasoning would make sense only if direct exciton recombination *via* CT is low compared to the charge separation. This property (illustrated in Fig. 6) is satisfied for all blends studied here since the recombination rates *via* CT, $k_{ER}$ and $k_{HR}$ (both between $10^5$ and $10^8$ s$^{-1}$), are lower than the charge separation rate, $k_{ES}$ and $k_{HS}$ (both between $10^9$ and $10^{10}$ s$^{-1}$). Using those arguments, a favorable condition for exciton dissociation is obtained when the variation of $\Delta G$ associated to charge separation is lower than the variation of $\Delta G$ for the charge back process, i.e., $\Delta G_{ES} < \Delta G_{EB}$ ($\Delta G_{HS} < \Delta G_{HB}$) for donor (acceptor) excitation. One can rewrite this relation using $\Delta G_{HS} = E_{b,CT}$ and $\Delta G_{HB} = -\Delta G_{HT}$ from Fig. 2c to find $E_{b,CT} < -\Delta G_{HT}$ or $E_{b,CT} + \Delta G_{HT} < 0$. Therefore, an efficient exciton dissociation is expected when the energy variation linked to charge transfer can compensate for the binding energy of the CT exciton. The same logic applies to derive the condition for an efficient electron transfer after the donor excitation ($E_{b,CT} + \Delta G_{ET} < 0$).

We then repeated the plots of Figs 7a and 7b but now using the criteria suggested in ref.[55]. The variation of PL quenching with $E_{b,CT} + \Delta G_{ET(HT)}$ is shown in Fig. 7c. It is clear that the curves for both $Q_D$ and $Q_A$ have similar behavior, but the results present the same asymmetry verified in Figs. 7a and b. $Q_D$ is shifted toward lower values of the parameter and again starts to fall before $Q_A$. Note that the second order polynomial fit of $Q_D$ and $Q_A$ is reasonable, but there is no single set of parameters that can describe both quenching simultaneously.

The reason behind the asymmetry between $Q_D$ and $Q_A$ in Figs 7a-c is the influence of FRET. For the blends in which D-A FRET is high, $Q_D$ is also determined by the driving forces involved in the hole transfer as well. Therefore, $Q_D$ is not fully defined only by the characteristic energies for the electron transfer. In Fig. 7d-f we replaced the experimental value of $Q_D$ by a theoretical quenching that is exclusively associated to the charge transfer (in other words, in the absence of FRET). This parameter was calculated from Eq. (3) assuming $p_{F,DA} = 0$. In those figures, we also plot the second-order polynomial fits of quenching efficiency. The previews asymmetry is removed and a much better simultaneous fit of those quantities, compared to the fittings in Figs 7a-c, is possible.

Comparing now Figs 7d and 7e with Fig 7f, one can see that the best simultaneous fit of $Q_D$ and $Q_A$ was found when the quenching is plotted as a function of $E_{b,CT} + \Delta G_{ET(HT)}$. This result suggests that the condition $E_{b,CT} + \Delta G_{ET(HT)} < 0$ can be a useful tool to anticipate the efficiency of the quenching process associated with the charge transfer. This is demonstrated in Fig 7f, where the highest values of $Q_D$ and $Q_A$ are indeed concentrated in the area that satisfies the relation $E_{b,CT} + \Delta G_{ET(HT)} < 0$.

The usefulness of the result in Fig. 7f comes from the fact that quenching ultimately depends on efficient charge transfer in either the presence or in the absence of FRET. D→A FRET only increases the weight of the exciton dissociation process *via* hole transfer (in the acceptor) over the exciton dissociation *via* electron transfer (in the donor). Under those circumstances, the maximization of $Q_A$ (by an efficient hole transfer) becomes even more crucial to the performance of OSCs.

Let us now explore an important consequence of our findings. From the definition of the driving force for hole transfer (Fig. 2) we have $\Delta G_{HT} = \Delta E_{HOMO} - (E_{b,A} - E_{b,CT})$. Assuming a negligible $\Delta E_{HOMO}$ (obtained, for example, by cyclic voltammetry), and an energy difference $E_{b,A} - E_{b,CT} \approx 0.1$ eV, one finds $\Delta G_{HT} = -0.1$ eV. Given this magnitude of $\Delta G_{HT}$, the only way to satisfy the condition $E_{b,CT} + \Delta G_{HT} \leq 0$ is by making $E_{b,CT} \leq 0.1$ eV, which implies that $E_{b,A} \leq 0.2$ eV. Therefore, to minimize driving force without decreasing the quenching efficiency, it is essential to simultaneously reduce both $E_{b,CT}$ and $E_{b,A}$. Additionally, it is also important to keep the difference between them as small as possible. Remember that to find the minimal driving force to promote the exciton dissociation is important to further increase the open-circuit voltage of OSCs. For this purpose, we can take advantage of the theoretical model and map the quenching efficiency by varying $E_{b,CT}$ and $\Delta G_{ET(HT)}$. It is important to mention that this prediction must be followed to optimize the process of exciton dissociation purely intermediated by the CT state *via* acceptor or donor excitation. The result presented in Fig. 8 makes clearer what was discussed above about the importance of the $E_{b,CT} + \Delta G_{ET(HT)} < 0$ (or $E_{b,CT} < -\Delta G_{ET(HT)}$) relationship. Below, we will go deeper into the influence of D→A FRET on quenching efficiency.

**3.3.2 The role of the Marcus Inverted Regime**

We will show that the strategy to enhance $Q_A$ by the maximization of the hole transfer tends also to increase the FRET probability $p_{F,DA}$ under some conditions. A subtle mechanism links the two parameters. Table 3 and Fig. 9 will help us to describe this relation. Table 3 compares $|\Delta G_{ET(HT)}|$ with the respective reorganization energies for electron (hole) transfer from D (A) to A (D). The Fig. 9 illustrates the schematic variation of $k_{ET(HT)}$, $p_{F,DA}$, and $Q_A$ with $E_{b,CT} + \Delta G_{HT}$ as this parameter gets more negative with decreasing $\Delta G_{ET}$. One can see that all blends are in the "Marcus inverted region" (MIR) for the electron injection characterized by $|\Delta G_{ET}| > \lambda_e^{D-A}$ (where $\lambda_e^{D-A}$ is the reorganization energy for the electron injection from D to A, see Table 3). As $E_{b,CT} + \Delta G_{HT}$ becomes more negative by the decrease of $\Delta G_{HT}$, $\Delta G_{ET}$ also tends to decrease due to the rigid shift of the acceptor gap. Due to the MIR, $k_{ET}$ will decrease with a decrease in $\Delta G_{ET}$. Since $k_{F,DA}$ almost does not change with $\Delta G_{ET(HT)}$, $p_{F,DA}$ increases following Eq. 5, enhancing the probability of energy transfer between D and A.

Contrary to electron transfer, hole injection from the acceptor to the donor can suffer a transition from the "Marcus direct region" (MDR) to the "Marcus inverted region" (MIR). When $|\Delta G_{HT}| < \lambda_h^{A-D}$ (MDR condition, where $\lambda_h^{A-D}$ is the reorganization energy for hole injection from the A to D), $k_{HT}$ and $Q_A$ increase with $\Delta G_{HT}$ becoming more negative. The systems IT-M, ITIC and ITCC are in this regime (see Fig. 9). For more negative values of $\Delta G_{HT}$, $k_{HT}$ reaches then a maximum at $|\Delta G_{HT}| = \lambda_h^{A-D}$, which tends to maximize $Q_A$. A further decrease of $\Delta G_{HT}$ will then induce a MDR to MIR transition in the interval $|\Delta G_{HT}| > \lambda_h^{A-D}$. The IT-nX systems are in this regime but with $|\Delta G_{HT}| \gtrsim \lambda_h^{A-D}$ so that $k_{HT}$ is close to the maximum and has a weak dependence on $\Delta G_{HT}$. This is exactly the reason why they have the best performance among the blends analyzed in this work (see Fig. 9). Yet a further decrease of $\Delta G_{HT}$ in the interval $|\Delta G_{HT}| > \lambda_h^{A-D}$ would decrease $Q_A$ (and $Q_D$) due to the reduction of the hole transfer.

The mechanism illustrated in Fig. 9 explains the relation between $p_{F,DA}$ and $Q_A$ in Table 2. It also suggests that the fulfillment of relations $E_{b,CT} + \Delta G_{HT} < 0$ and $|\Delta G_{HT}| \approx \lambda_h^{A-D}$ are sufficient general to serve as a guideline to the optimization of the OSC performance in the presence of FRET. Thinking about energy loss reduction, we conclude that $\Delta G_{HT}$ minimization must be accompanied by $\lambda_h^{A-D}$ minimization to maintain the maximum hole transfer rate and quenching efficiency. Indeed, a consequence of those conditions is that the systems with $E_{b,CT} < \lambda_h^{A-D}$ tend to have the highest $Q_D$ and $Q_A$. This can be verified by inspecting the 5$^{th}$ column of Table 3.

### 3.3.3 Singlet exciton lifetime

Recently, the great importance of acceptor singlet exciton lifetime for free charge generation at negligible energy-level offsets was verified by Classen and coworkers.[105] They found that the high efficient molecular acceptor Y6[106] has a very long exciton lifetime of 1016 ps, which enables efficient exciton dissociation at small energy-level offsets. From our kinetic model it is possible to easily investigate the influence of singlet exciton recombination lifetime in $Q_A$.

In Fig. 9 we present the results of $Q_A$ for different singlet exciton recombination lifetimes. It is observed that indeed this parameter has a strong influence on the magnitudes of the quenching efficiency. Fig. 9 again demonstrate that the condition $E_{b,CT} + \Delta G_{HT} < 0$ establishes a limit for efficient exciton quenching. In the range of values where this condition is violated, the quenching efficiency quickly tends to zero. Yet the highest magnitudes of those parameters in the $E_{b,CT} + \Delta G_{HT} < 0$ interval depend on exciton-lifetime in the acceptor phases as well. These results indicate that it is possible to use a simple relation based on the exciton binding energy and the driving force for charge transfer as a criterion to anticipate the quenching efficiency. Yet, those results also show that this criterion is necessary but not sufficient to maximize exciton dissociation. Besides the fulfillment of the proposed relation, a longer singlet exciton lifetime in the donor and acceptor is also fundamental to enable an efficient free charge generation in D/A blends.

## 4. Conclusions

In this work we generalize a previously reported kinetic model, including the FRET mechanism, to reproduce the efficiency of PL quenching reported in the literature for ten D/A blends. Those blends preset different driving forces for electron and hole transfer. An important parameter of the model is the exciton binding energy, therefore, we developed a careful theoretical approach to the determination of the film's dielectric constant, which enables us to estimate the influence of electronic polarization effects on exciton binding energy of acceptor molecules. From these results, it was possible to determine the driving forces and the transition rates using the Marcus/Hush theory combined with disorder considerations. The FRET rates were also calculated considering the spectral overlap of the PBDB-TF emission and NFA absorption. We first found that the model excellently reproduces the experimental quenching efficiency for acceptor excitation for all blends after a small adjustment of $E_{b,CT}$ from $E_{b,A}$ (no more than 0.1 eV was necessary). For donor excitation, the theoretical quenching efficiency reproduces the experimental results exceptionally evidencing the great relevance of the FRET mechanism for some blends. The donor-acceptor FRET followed by hole transfer processes puts the exciton dissociation process via hole transfer at a higher level compared to the electron transfer.

Motivated by the good correlation between experiment and theory, we then explored the variations of PL quenching with some important parameters of the model, like the driving forces, the exciton binding energy and exciton lifetime. The idea was to search for criteria that would characterize systems with high quenching efficiency. The first important point is that an efficient quenching is observed when the binding energy of the CT state is lower than the negative of the driving force for hole or electron transfer. Specifically, when a D/A blend simultaneously satisfy the relations $E_{b,CT} < -\Delta G_{ET}$ and $E_{b,CT} < -\Delta G_{HT}$, it tends to have an equally efficient PL quenching for photoexcitation of either the donor or acceptor phases. This was observed for the blends containing the PBDB-TF polymer with the halogenated derivatives of ITIC. For the other blends that satisfies this criterion only for the electron transfer so that $E_{b,CT} < -\Delta G_{ET}$ (but $E_{b,CT} > -\Delta G_{HT}$) the lower $Q_A$ can affect the high $Q_D$ depending on the D-A FRET probability. In the presence of FRET, $Q_D$ is not completely determined by the characteristic energies of the electron transfer alone, but also depends on the driving forces involved in the hole transfer as well. In this case, the fulfillment of relations $E_{b,CT} < -\Delta G_{HT}$ and $|\Delta G_{HT}| \approx \lambda_h^{A-D}$ is sufficient general to serve as a guideline to the optimization of OSC performance. The strategy of $\Delta G_{HT}$ minimization must necessarily be accompanied by the minimization of the reorganization energy for hole transfer from A to D. This procedure would maintain the maximum hole transfer rate, which will enhance the quenching efficiency.

We also observed that when the proposed condition for high quenching is valid, systems with longer exciton lifetimes will have maximum quenching efficiency that approaches 100%. As the exciton lifetime becomes shorter, the highest values of quenching decreases. Finally, we conclude that to reconcile low driving force with high quenching efficiency it is necessary that the binding energy of the CT exciton and the local exciton are at the same time low and close to each other. Our results offer a comprehensive framework to understand the exciton dissociation process in D/A blends and propose a rule of thumb principle to guide the development of more efficient OSCs based on D/A blends.

**Conflicts of interest**

The authors declare no conflicts of interest.

**Acknowledgments**

This work has been supported by the Companhia Paranaense de Energia – COPEL research and technological development program, through the PD 2866-0470/2017 project, regulated by ANEEL. This study was financed in part by the Coordenação de Aperfeiçoamento de Pessoal de Nível Superior-Brasil (CAPES)-Finance Code 001. Research developed with the assistance of CENAPAD-SP (Centro Nacional de Processamento de Alto Desempenho em São Paulo), project UNICAMP/FINEP – MCT and Sistema Nacional de Processamento de Alto Desempenho (SINAPAD/SDUMONT). Special thanks go to CNPq (grant 381113/2021-3) and to LCNano/SisNANO 2.0 (grant 442591/2019-5) for financial support. G. Candiotto gratefully acknowledges FAPERJ Process E-26/200.008/2020 for financial support and Núcleo Avançado de Computação de Alto Desempenho (NACAD/COPPE/UFRJ).

**FIGURES**

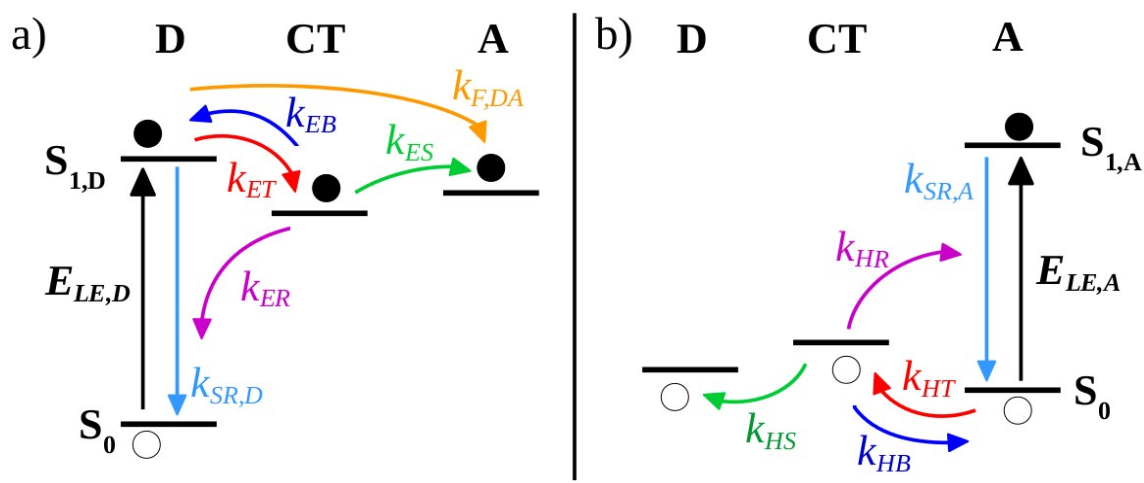

**Figure 1** – Simplified charge dynamics diagram at the D/A interface after (a) donor excitation (electron dynamics) and (b) acceptor excitation (hole dynamics).

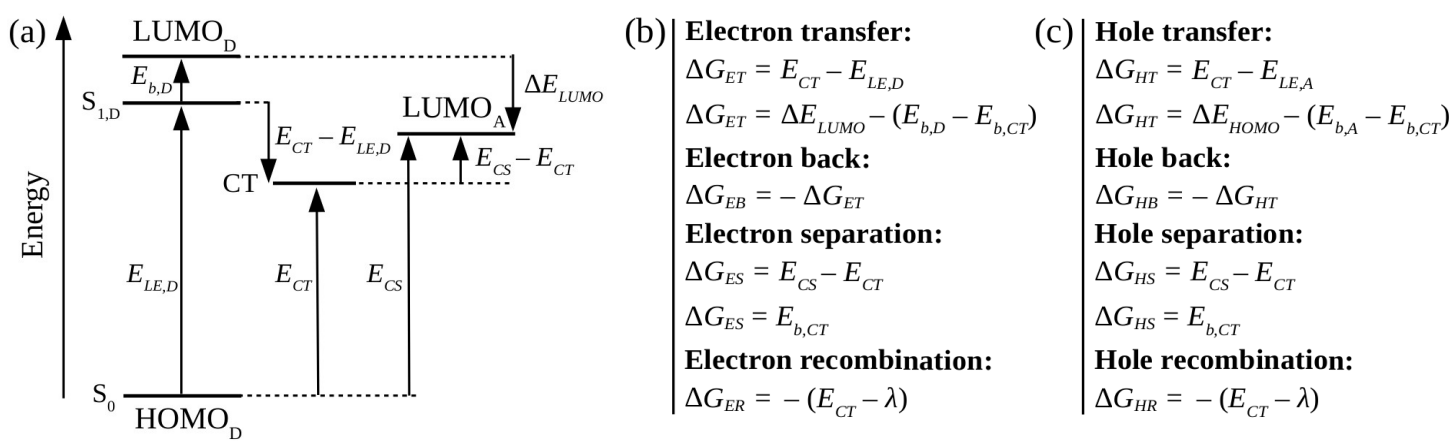

**Figure 2** – (a) Schematic of the state energy. Driving forces for electron dynamics in (b) and hole dynamics in (c).

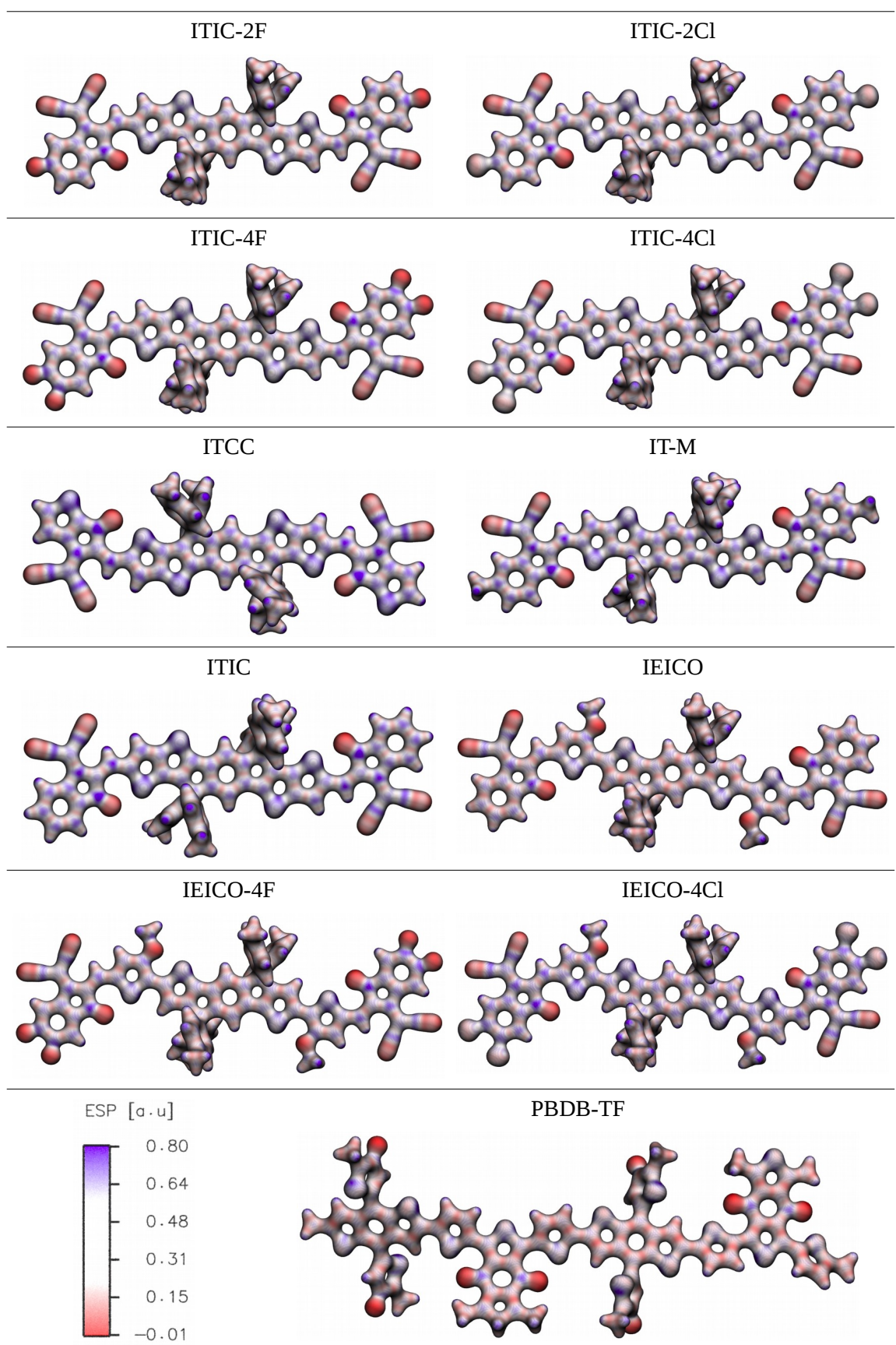

**Figure 3** - ESP distributions from molecular acceptors and PBDB-TF donor polymer with two mers.

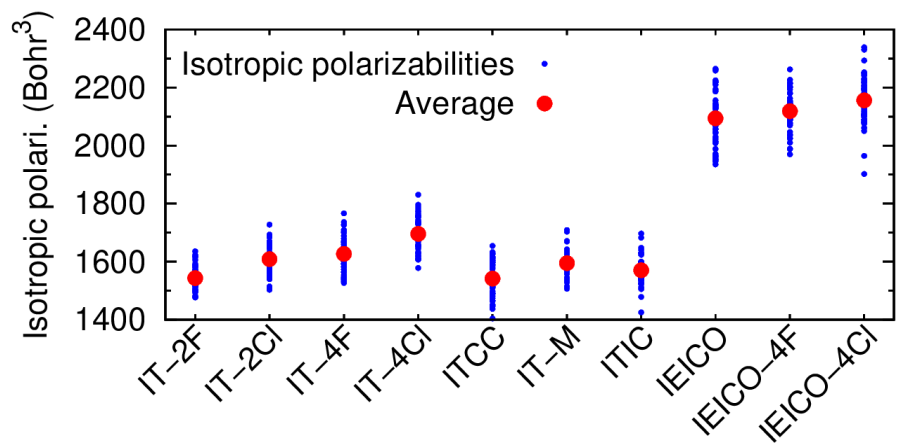

**Figure 4** - Isotropic polarizability for each of the 50 acceptors extracted from the molecular cluster and the average value. Each value was obtained *via* DFT considering individual molecules in the final geometry of MD.

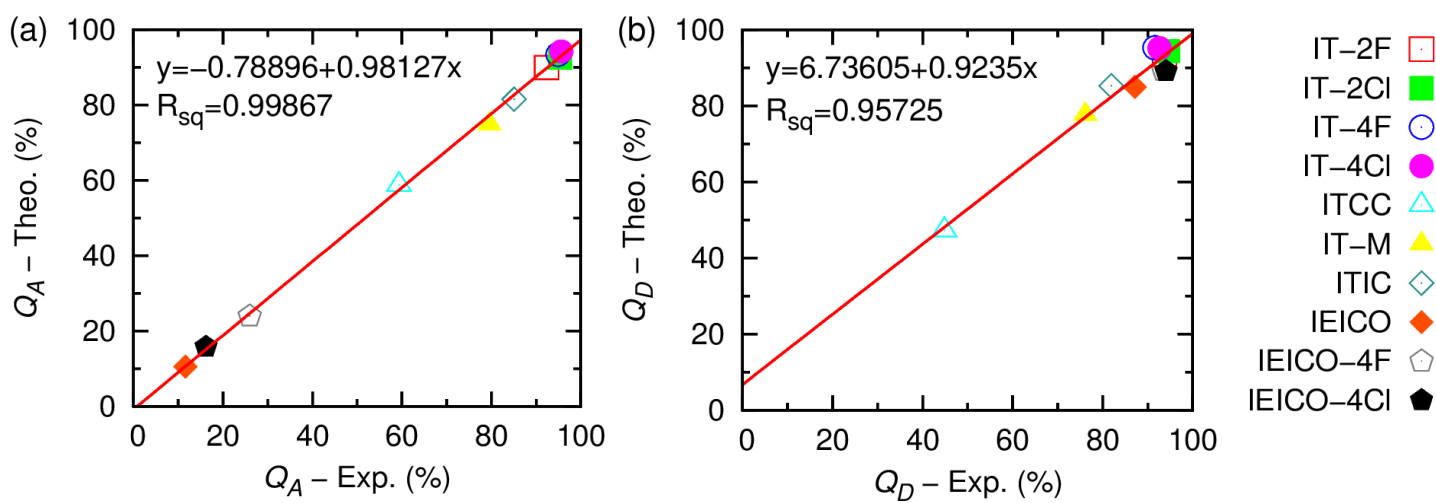

**Figure 5** - Correlation between theoretical and experimental results of (a) $Q_A$ and (b) $Q_D$ for the PBDB-TF-based blends. The linear equation and the $R_{sq}$ are displayed in detail.

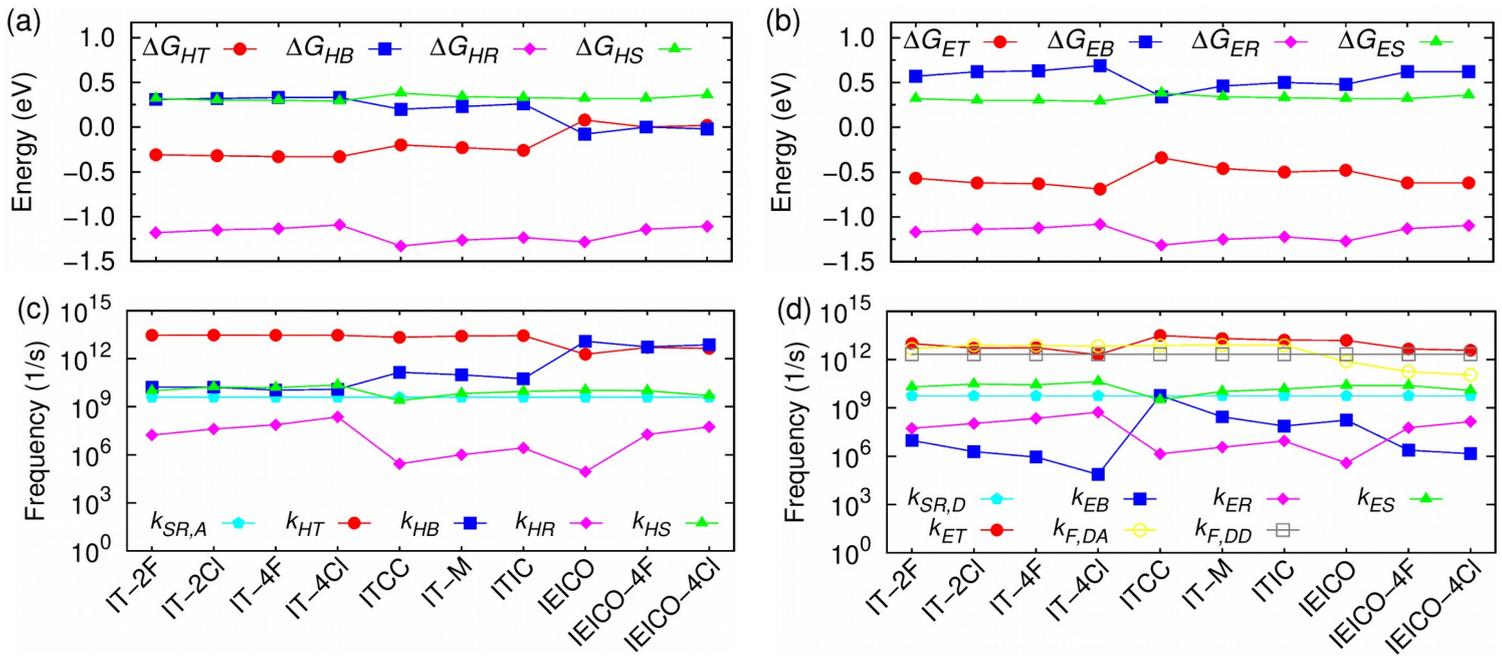

**Figure 6** – Driving forces of the PBDB-TF-based blends for the electron (a) and hole (b) dynamics. Averaged rates for the electron (c) and hole (d) dynamics. The singlet exciton recombination lifetime of 178 ps for PBDB-T (from Ref.[10]) and 256 ps for ITIC (from Ref.[82]) was used to obtain $k_{SR,D}$ = 5.62 x $10^9$ s$^{-1}$ and $k_{SR,A}$ = 3.91 x $10^9$ s$^{-1}$.

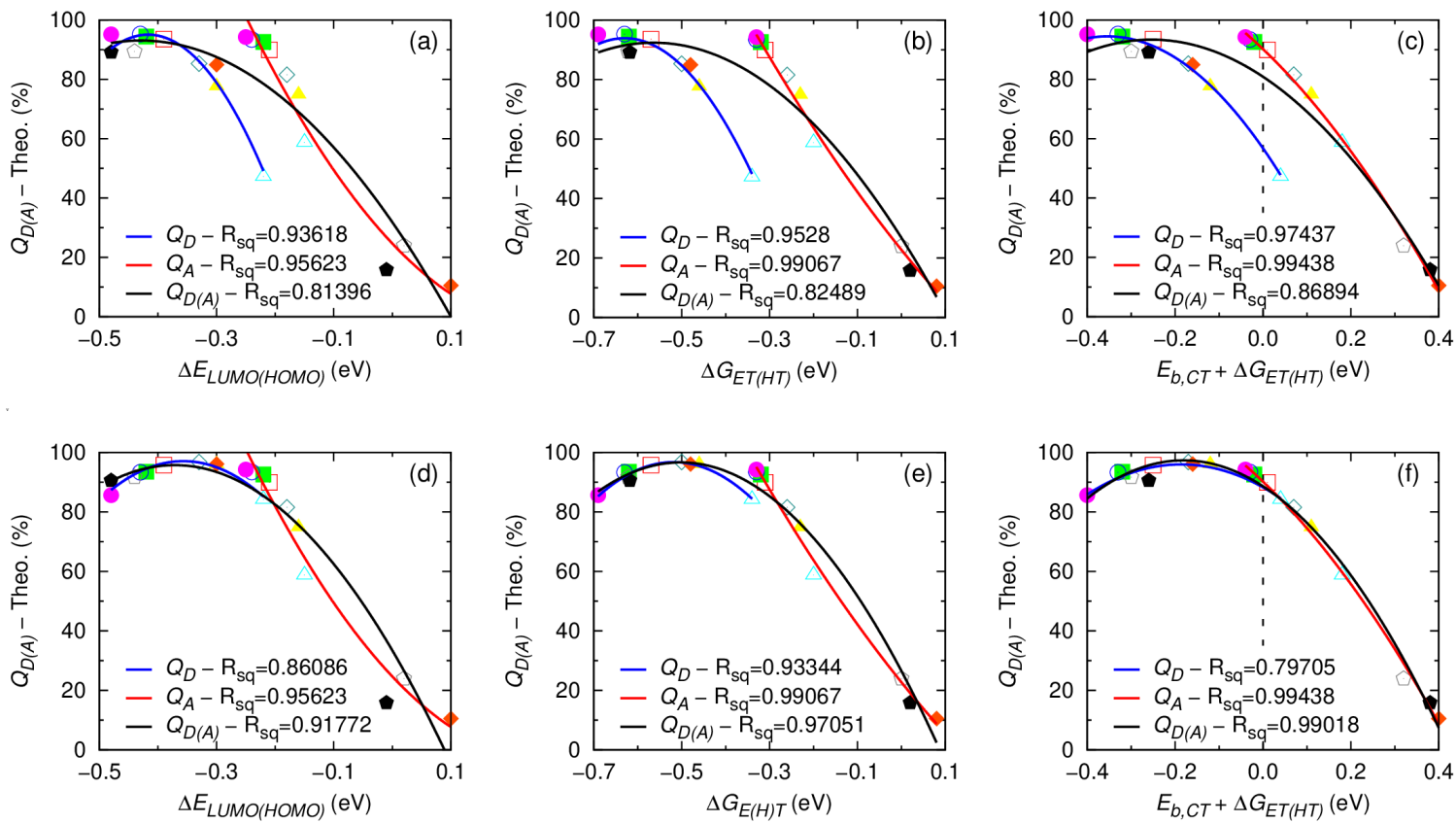

**Figure 7** – Quenching efficiency and state energies. The lines are least square fit of a second order polynomial, to the data points, where the $R_{sq}$ are displayed in detail. The $Q_D$ results in (a), (b) and (c) were obtained with equation (3) that consider FRET. The $Q_D$ results in (d), (e) and (f) were obtained with equation (3) but not considering FRET, by setting $p_{F,DA} = 0$. The vertical dashed black line in (c) and (f) specifies the zero of X-axes.

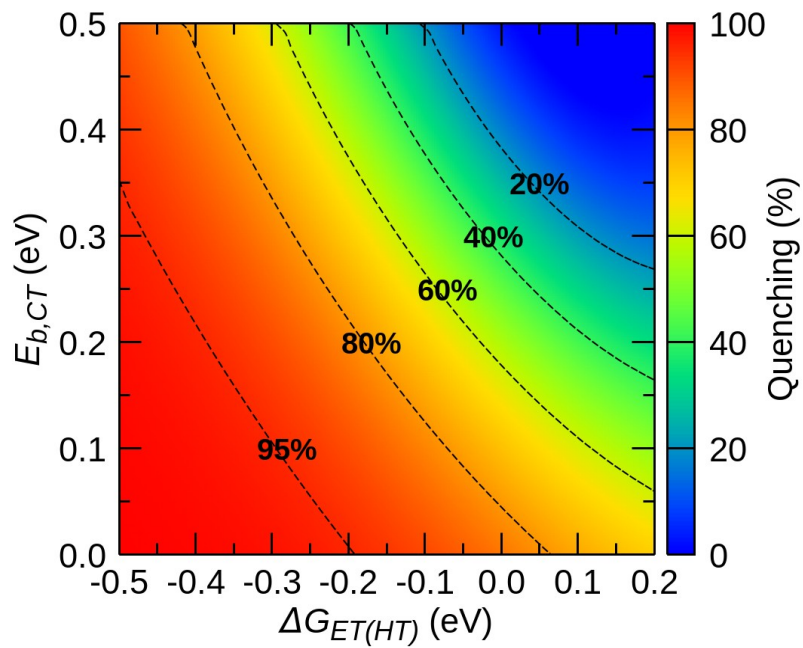

**Figure 8** – Theoretical prediction of photoluminescence quenching efficiency purely intermediated by CT state by varying the driving force for CT formation and binding energy of CT state.

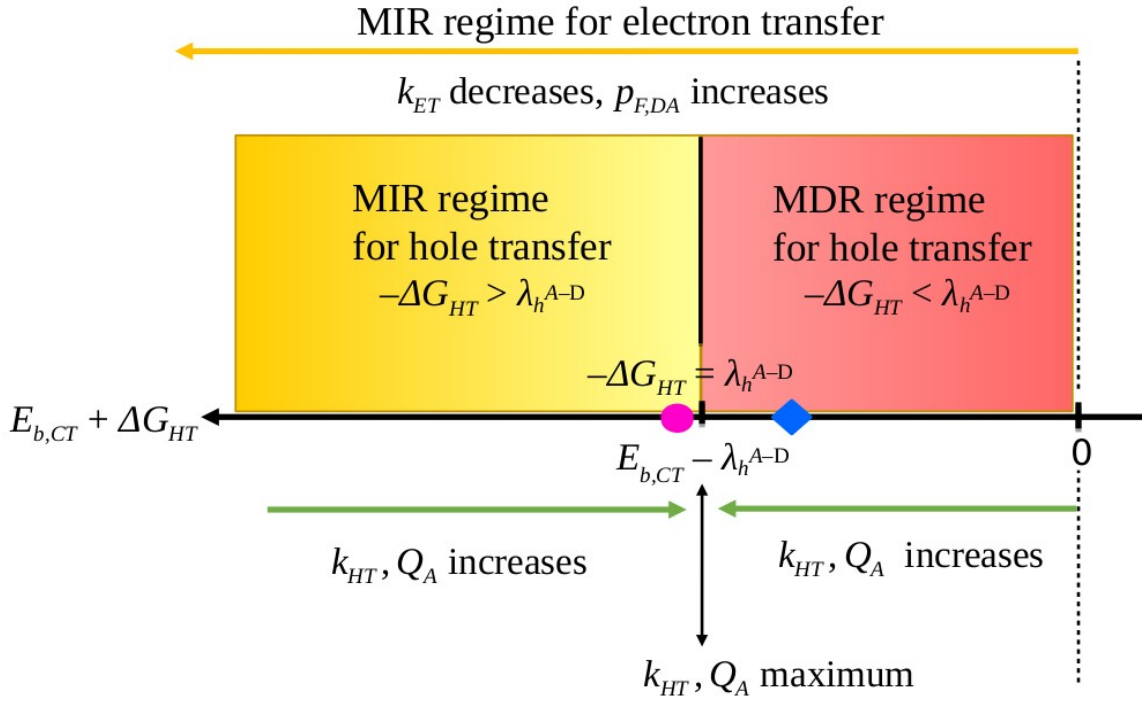

**Figure 9** –Schematic variation of $k_{ET(HT)}$, $p_{F,DA}$, and $Q_A$ with $E_{b,CT} + \Delta G_{HT}$ as this parameter gets more negative with decreasing $\Delta G_{HT}$. The circle indicates the schematic localization of the IT-nX system whereas the diamond indicates the schematic localization of the ITIC and IT-M systems.

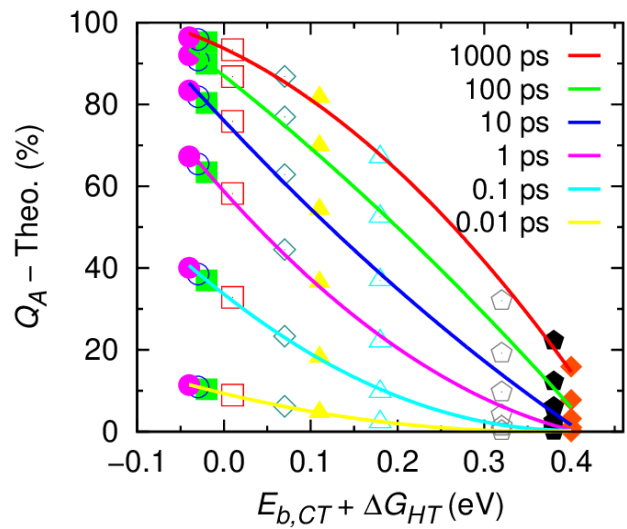

**Figure 10** - Quenching efficiency for acceptor excitation and state energies. Results for different singlet exciton lifetimes in acceptor. The lines are least square fit of a second order polynomial, to the data points, where $R_{sq}$ was omitted because it is greater than 0.99 for all curves.

# TABLES

**Table 1** - Molecular volume, mean electronic polarizability, dielectric constant, gas-phase exciton binding energy, solid-state exciton binding energy and exciton binding energy of charge transfer state.

| Materials | $V_M$ (Bohr$^3$/mol) | Mean $\alpha$ (Bohr$^3$) | $\varepsilon$ | $E_b^{vac}$ (eV) | $E_b^{vac}/\varepsilon$ (eV) | $E_{b,CT}$ (eV) |
|---|---|---|---|---|---|---|
| IT-2F | 12241.18 | 1543.62 | 4.36 | 1.84 | 0.42 | 0.32 |
| IT-2Cl | 12394.67 | 1608.88 | 4.57 | 1.82 | 0.40 | 0.30 |
| IT-4F | 12279.55 | 1626.56 | 4.74 | 1.84 | 0.39 | 0.30 |
| IT-4Cl | 12586.54 | 1695.75 | 4.89 | 1.80 | 0.37 | 0.29 |
| ITCC | 11972.56 | 1541.56 | 4.51 | 1.95 | 0.43 | 0.38 |
| IT-M | 12471.42 | 1595.18 | 4.46 | 1.84 | 0.41 | 0.34 |
| ITIC | 12164.43 | 1570.48 | 4.53 (4.50$^a$) | 1.85 | 0.41 | 0.33 |
| IEICO | 15426.19 | 2094.16 | 4.95 | 1.69 | 0.34 | 0.32 |
| IEICO-4F | 15541.31 | 2118.45 | 4.99 | 1.68 | 0.34 | 0.32 |
| IEICO-4Cl | 15848.30 | 2155.94 | 4.97 | 1.64 | 0.33 | 0.36 |
| PBDB-TF | - | - | 4.00$^b$ | 2.01 | 0.50 | - |

$^a$Obtained from capacitance–voltage measurements by ref.[107]. $^b$A dielectric constant of 4.0 was used in the calculations in accordance with other works.[56,70,71]

**Table 2** – The probability of D→A FRET ($p_{F,DA}$), the ratio of electron ($\eta_{CT}$) and FRET-hole ($\eta_F$) processes in the total quenching loss, and the quenching efficiency upon acceptor illumination ($Q_A$). Results in %.

| Blends | $p_{F,DA}$ | $\eta_{CT}$ | $\eta_F$ | $Q_A$ |
|---|---|---|---|---|
| D/IT-2F | 44.6 | 0.9 | 99.1 | 92.34 |
| D/IT-2Cl | 61.2 | 0.4 | 99.6 | 95.37 |
| D/IT-4F | 59.7 | 0.6 | 99.4 | 94.9 |
| D/IT-4Cl | 67.5 | 0.6 | 99.4 | 95.6 |
| D/ITCC | 30.3 | 0.2 | 99.8 | 59.3 |
| D/IT-M | 38.6 | 0.2 | 99.8 | 79.6 |
| D/ITIC | 43.7 | 0.3 | 99.7 | 85.1 |
| D/IEICO | 9.1 | 38.0 | 62.0 | 11.6 |
| D/IEICO-4F | 4.9 | 36.6 | 63.4 | 26.0 |
| D/IEICO-4Cl | 3.0 | 55.0 | 45.0 | 16.2 |

**Table 3** – The driving forces and reorganization energies for electron and hole transfer. In yellow are the processes that are in the Marcus inverted region (MIR) whereas in red are the processes in the Marcus direct region (MDR). The last column is the difference between the exciton binding energy and the reorganization energy for hole transfer. The two colors of this column differentiate the systems with negative and positive values of this parameter. Results in eV.

| Blends | $|\Delta G_{HT}|$ | $\lambda_h^{A-D}$ | $|\Delta G_{ET}|$ | $\lambda_e^{D-A}$ | $E_{b,CT} - \lambda_h^{A-D}$ |
|---|---|---|---|---|---|
| D/IT-2F | 0.31 | 0.309 | 0.57 | 0.321 | -0.011 |
| D/IT-2Cl | 0.32 | 0.310 | 0.62 | 0.322 | -0.01 |
| D/IT-4F | 0.33 | 0.315 | 0.63 | 0.327 | -0.015 |
| D/IT-4Cl | 0.33 | 0.307 | 0.69 | 0.319 | -0.017 |
| D/ITCC | 0.20 | 0.330 | 0.34 | 0.345 | 0.05 |
| D/IT-M | 0.23 | 0.317 | 0.46 | 0.329 | 0.023 |
| D/ITIC | 0.20 | 0.315 | 0.50 | 0.327 | 0.015 |
| D/IEICO | 0.08 | 0.296 | 0.48 | 0.309 | 0.024 |
| D/IEICO-4F | 0.00 | 0.297 | 0.62 | 0.309 | 0.023 |
| D/IEICO-4Cl | 0.26 | 0.291 | 0.62 | 0.303 | 0.069 |